%% file: main.tex
\documentclass[runningheads]{llncs}
\usepackage[T1]{fontenc}
\usepackage{xcolor}
\usepackage{amsfonts}
\usepackage{amsmath}
\usepackage{booktabs}
\usepackage[para]{footmisc}
\usepackage{soul}

\setstcolor{red}
\definecolor{rel0}{HTML}{eb4034}
\definecolor{rel1}{HTML}{fcba03}
\definecolor{rel2}{HTML}{4287f5}
\definecolor{rel3}{HTML}{32a852}

\newcommand{\crcedit}[1]{#1}

\usepackage{todonotes}
\usepackage{graphicx}
\begin{document}
\title{Doc2Query\--\--: When Less is More}
\author{Mitko Gospodinov\inst{1} \and Sean MacAvaney\inst{2} \and Craig Macdonald\inst{2}}
\institute{University of Glasgow\\
\inst{1}\email{2024810G@student.gla.ac.uk}\\
\inst{2}\email{\{first\}.\{last\}@glasgow.ac.uk}
}
\authorrunning{Gospodinov et al.}
\maketitle              %
\begin{abstract}
Doc2Query --- the process of expanding the content of a document before indexing using a sequence-to-sequence model --- has emerged as a prominent technique for improving the first-stage retrieval effectiveness of search engines. However, sequence-to-sequence models are known to be prone to ``hallucinating'' content that is not present in the source text. We argue that Doc2Query is indeed prone to hallucination, which ultimately harms retrieval effectiveness and inflates the index size. In this work, we explore techniques for filtering out these harmful queries prior to indexing. We find that using a relevance model to remove poor-\crcedit{quality} queries can improve the retrieval effectiveness of Doc2Query by up to 16\%, while simultaneously reducing mean query execution time by 23\% and cutting the index size by 33\%. \crcedit{We release the code, data, and a live demonstration to facilitate reproduction and further exploration.}\footnote{\url{https://github.com/terrierteam/pyterrier\_doc2query}\label{fn:code}}

\end{abstract}

\section{Introduction}

Neural network models, particularly those based on contextualised language models, have been shown to improve search effectiveness~\cite{bertimprove}. While some approaches focus on re-ranking document sets from a first-stage retrieval function to improve precision~\cite{bertrerank}, others aim to improve the first stage itself~\cite{hdct}. In this work, we focus on one of these first-stage approaches: Doc2Query~\cite{original-d2q}. This approach trains a sequence-to-sequence model (e.g., T5~\cite{t5}) to predict queries that may be relevant to a particular text. Then, when indexing, this model is used to \textit{expand} the document by generating a collection of queries and appending them to the document. Though computationally expensive at index time~\cite{carbon}, this approach has been shown to be remarkably effective even \crcedit{when} retrieving using \crcedit{simple} lexical models like BM25~\cite{doct5query}. Numerous works have shown that the approach can produce a high-quality pool of results that are effective for subsequent stages in the ranking pipeline~\cite{epic,gar,deepimpact,tilde}.

However, sequence-to-sequence models are well-known to be prone to generate content that does not reflect the input text -- a defect known in literature as ``hallucination''~\cite{maynez-etal-2020-faithfulness}. We find that existing Doc2Query models are no exception. Figure~\ref{fig:ex_msm} provides example generated queries from the state-of-the-art T5 Doc2Query model~\cite{doct5query}. In this example, we see that \crcedit{many of the} generated queries cannot actually be answered by the source passage (score $\leq$ 1).

\input{fig_ex_msm}

Based on this observation, we hypothesise that retrieval performance of Doc2Query would improve if hallucinated queries were removed. In this paper, we conduct experiments where we apply a new filtering phase that aims to remove poor queries prior to indexing. Given that this approach \textit{removes} queries, we call the approach Doc2Query\--\-- (Doc2Query-minus-minus). Rather than training a new model for this task, we identify that relevance models are already fit for this purpose: they estimate how relevant a passage is to a query. We therefore explore filtering strategies that make use of existing neural relevance models.

Through experimentation on the MS MARCO dataset, we find that our filtering approach can improve the retrieval effectiveness of indexes built using Doc2Query\--\-- by up to 16\%; less can indeed be more. Meanwhile, filtering naturally reduces the index size, lowering storage and query-time computational costs. Finally, we conduct an exploration of the index-time overheads introduced by the filtering process and conclude that the gains from filtering more than make up for the additional time spent generating more queries. \crcedit{The approach also has a positive impact on the environmental costs of applying Doc2Query; the same retrieval effectiveness can be achieved with only about a third of the computational cost when indexing.} \crcedit{To facilitate last-metre, last-mile, and complete reproduction efforts~\cite{10.1145/3477495.3531721}, we release the code, indices, and filtering scores.\footref{fn:code}} In summary, we contribute a technique to improve the effectiveness and efficiency of Doc2Query by filtering out queries that do not reflect the original passage.

\section{Related Work}\label{s:related}

The classical lexical mismatch problem is a key one in information retrieval - documents that do not contain the query terms may not be retrieved. In the literature, various approaches have addressed this: query reformulation  -- including stemming, query expansion models (e.g.\, Rocchio, Bo1~\cite{bo1}, RM3~\cite{rm3}) -- and document expansion~\cite{10.1145/2348283.2348405,10.1145/1871437.1871571,10.3115/1220835.1220887}. Classically, query expansion models have been popular, as they avoid the costs associated with making additional processing for each document needed for document expansion. However, query expansion may result in reduced performance~\cite{query_expansion}, as queries are typically short and the necessary evidence to understand the context of the user is limited. 

\looseness -1 The application of latent representations of queries and documents, such as using latent semantic indexing~\cite{latent_indexing} allow retrieval to not be driven directly by lexical signals. More recently, transformer-based language models (such as BERT~\cite{bert}) have resulted in representations of text where the contextualised meaning of words are accounted for. In particular, in dense retrieval, queries and documents are represented in embeddings spaces~\cite{colbert,ance}, often facilitated by Approximate Nearest Neighbour (ANN) data structures~\cite{faiss}. However, even when using ANN, retrieval can still be inefficient or insufficiently effective~\cite{Lin2021OnTS}.

Others have explored approaches for augmenting lexical representations with additional terms that may be relevant. \crcedit{In this work, we explore Doc2Query~\cite{original-d2q}, which uses a sequence-to-sequence model that maps a document to queries that it might be able to answer. By appending these {\em generated} queries to a document's content before indexing, the document is more likely to be retrieved for user queries when using a model like BM25. An alternative style of document expansion, proposed by MacAvaney~et~al.~\cite{epic} and since used by several other models (e.g.,~\cite{splade,sparta,tilde}), uses the built-in Masked Language Modelling (MLM) mechanism. MLM expansion generates individual tokens to append to the document as a bag of words (rather than as a sequence). Although MLM expansion is also prone to hallucination,\footnote{\crcedit{For instance, we find that SPLADE~\cite{splade} generates the following seemingly-unrelated terms for the passage in Figure~\ref{fig:ex_msm} in the top 20 expansion terms: \textit{reed}, \textit{herb}, and \textit{troy}.}} the bag-of-words nature of MLM expansion means that individual expansion tokens may not have sufficient context to apply filtering effectively. We therefore focus only on sequence-style expansion and leave the exploration of MLM expansion for future work.}

\section{Doc2Query\--\--}\label{s:method}

Doc2Query\--\-- consists of two phases: a \textit{generation} phrase and a \textit{filtering} phase. In the generation phase, a Doc2Query model generates a set of $n$ queries that each document might be able to answer. However, as shown in Figure~\ref{fig:ex_msm}, not all of the queries are necessarily relevant to the document. To mitigate this problem, Doc2Query\--\-- then proceeds to a filtering phase, which is responsible for eliminating the generated queries that are least relevant to the source document. Because hallucinated queries contain details not present in the original text (by definition), we argue that hallucinated queries are less \crcedit{useful for retrieval} than non-hallucinated ones. Filtering is accomplished by retaining only the most relevant $p$ proportion of generated queries over the entire corpus. The retained queries are then concatenated to their corresponding documents prior to indexing, as per the existing Doc2Query approach.

More formally, consider an expansion function $e$ that maps a document to $n$ queries: $e:\textbf{D}\mapsto \textbf{Q}^n$. In Doc2Query, each document in corpus $\mathcal{D}$ are concatenated with their expansion queries, forming a new corpus  $\mathcal{D}^\prime=\{\text{Concat}(d, e(d)) \mid d \in \mathcal{D}\}$, which is then indexed by a retrieval system. Doc2Query\--\-- adds a filtering mechanism that uses a relevance \crcedit{model} that maps a query and document to a real-valued \crcedit{relevance score} $s:\textbf{Q} \times \textbf{D}\mapsto \mathbb{R}$ (with larger values indicating higher relevance). The relevance scoring function is used to filter down the queries to those that meet a certain score threshold $t$ as follows:

\begin{equation}
\mathcal{D}^\prime=\Big\{\text{Concat}\big(d, \big\{q \mid q \in e(d) \land s(q,d) \geq t\big\}\big) \mid d \in \mathcal{D}\Big\}
\end{equation}

The relevance threshold $t$ is naturally dependent upon the relevance scoring function. It can be set empirically, chosen based on operational criteria (e.g., target index size), or (for a well-calibrated relevance scoring function) determined \textit{a priori}. In this work, we combine the first two strategies: we pick $t$ based on the distribution of relevance scores across all expansion queries. \crcedit{For instance, at $p=0.3$ we only keep queries with relevance scores in the top 30\%, which is $t=3.215$ for the ELECTRA~\cite{electra} scoring model on the MS MARCO dataset~\cite{msmarco}.}

\section{Experimental Setup}\label{s:exp}

We conduct experiments to answer the following research questions:
\begin{enumerate}
\item[RQ1] Does Doc2Query\--\-- improve the effectiveness of document expansion?
\item[RQ2] What are the trade-offs in terms of effectiveness, efficiency, and storage when using Doc2Query\--\--?
\end{enumerate}

\textbf{Datasets and Measures.} We conduct tests using the MS MARCO~\cite{msmarco} v1 passage corpus. We use five test collections:\footnote{ir-datasets~\cite{macavaney:sigir2021-irds} IDs: \texttt{msmarco-passage/dev/small}, \texttt{msmarco-passage/dev/2}, \texttt{msmarco-passage/eval/small}, \texttt{msmarco-passage/trec-dl-2019/judged}, \texttt{msmarco-passage/trec-dl-2020/judged}} (1) the MS MARCO Dev (small) collection, consisting of 6,980 queries (1.1 qrels/query); (2) the Dev2 collection, consisting of 4,281 (1.1 qrels/query); (3) the MS MARCO Eval set, consisting of 6,837 queries (held-out leaderboard set); (4/5) the TREC DL'19/'20 collections, consisting of 43/54 queries (215/211 qrels/query). We evaluate using the official task evaluation measures: Reciprocal Rank at 10 (RR@10) for Dev/Dev2/Eval, nDCG@10 for DL'19/'20. We tune systems\footnote{BM25's $k1$, $b$, and whether to remove stopwords were tuned for all systems; the filtering percentage ($p$) was also tuned for filtered systems.} on Dev, leaving the remaining collections as held-out test sets.

\textbf{Models.} We use the T5 Doc2Query model from \crcedit{Nogueira and Lin}~\cite{doct5query}, making use of the inferred queries released by the authors (80 per passage). To the best of our knowledge, this is the highest-performing Doc2Query model available. We consider three neural relevance models for filtering: ELECTRA\footnote{\texttt{crystina-z/monoELECTRA\_LCE\_nneg31}}~\cite{electra}, MonoT5\footnote{\texttt{castorini/monot5-base-msmarco}}~\cite{monot5}, and TCT-ColBERT\footnote{\texttt{castorini/tct\_colbert-v2-hnp-msmarco}}~\cite{tct-colbert}, covering two strong cross-encoder models and one strong bi-encoder model. 
We also explored filters that use the probabilities from the generation process itself but found them to be ineffective and therefore omit these results due to space constraints.

\textbf{Tools and Environment.} We use the PyTerrier toolkit~\cite{pyterrier} with a PISA~\cite{pisa,pt_pisa} index to conduct our experiments. \crcedit{We deploy PISA's  Block-Max WAND~\cite{ding2011faster} implementation for BM25 retrieval.} Inference was conducted on an NVIDIA 3090 GPU. Evaluation was conducted using the ir-measures package~\cite{irm}.

\setlength{\tabcolsep}{4pt}

\begin{table}[]
\centering
\caption{Effectiveness and efficiency measurements for Doc2Query\--\-- and baselines. Significant differences between Doc2Query and their corresponding filtered versions for Dev, Dev2, DL'19 and DL'20 are indicated with * (paired t-test, $p<0.05$). Values marked with $\dagger$ are taken from the corresponding submissions to the public leaderboard.}
\label{tab:main}
\begin{tabular}{lrrrrrrr}
&\multicolumn{3}{c}{RR@10}&\multicolumn{2}{c}{nDCG@10}&\multicolumn{1}{c}{ms/q}&\multicolumn{1}{c}{GB}\\
\cmidrule(lr){2-4}\cmidrule(lr){5-6}\cmidrule(lr){7-7}\cmidrule(lr){8-8}
System & Dev & Dev2 & Eval & DL'19 & DL'20 & MRT & Index \\
\toprule

BM25 & 0.185 & 0.182 & $^\dagger$0.186 & 0.499 & 0.479 & 5 & 0.71 \\
\midrule
Doc2Query ($n=40$) & 0.277 & 0.265 & $^\dagger$0.272 & 0.626 & 0.607 & 30 & 1.17 \\
w/ ELECTRA Filter (30\%) &\bf *0.316 &\bf *0.310 & - &\bf 0.667 &\bf 0.611 &\bf 23 &\bf 0.89 \\
w/ MonoT5 Filter (40\%) & *0.308 & *0.298 & 0.306 & 0.650 &\bf 0.611 & 29 & 0.93 \\
w/ TCT Filter (50\%) & *0.287 & *0.280 & - & 0.640 & 0.599 & 30 & 0.94 \\
\midrule
Doc2Query ($n=80$) & 0.279 & 0.267 & - & 0.627 & 0.605 & 30 & 1.41 \\
w/ ELECTRA Filter (30\%) &\bf *0.323 &\bf *0.316 & 0.325 &\bf 0.670 &\bf 0.614 &\bf 23 &\bf 0.95 \\
w/ MonoT5 Filter (40\%) & *0.311 & *0.298 & - & 0.665 & 0.609 & 28 & 1.04 \\
w/ TCT Filter (50\%) & *0.293 & *0.283 & - & 0.642 & 0.588 & 28 & 1.05 \\

\bottomrule
\end{tabular}
\end{table}

\section{Results}\label{s:res}

We first explore RQ1: whether relevance filtering can improve the retrieval of Doc2Query models. Table~\ref{tab:main} compares the effectiveness of Doc2Query with various filters. We observe that all the filters significantly improve the retrieval effectiveness on the Dev and Dev2 datasets at both $n=40$ and $n=80$. We also observe a large boost in performance on the Eval dataset.\footnote{Significance cannot be determined due to the held-out nature of the dataset. Further, due to restrictions on the number of submissions to the leaderboard, we only are able to submit two runs. The first aims to be a fair comparison with the existing Doc2Query Eval result, using the same number of generated queries and same base T5 model for scoring. The second is our overall best-performing setting, using the ELECTRA filter at $n=80$ generated queries.} Though the differences in DL'19 and DL'20 appear to be considerable (e.g., 0.627 to 0.670), these differences are not statistically significant.

\begin{figure}[t]
    \centering
    \includegraphics[scale=0.68]{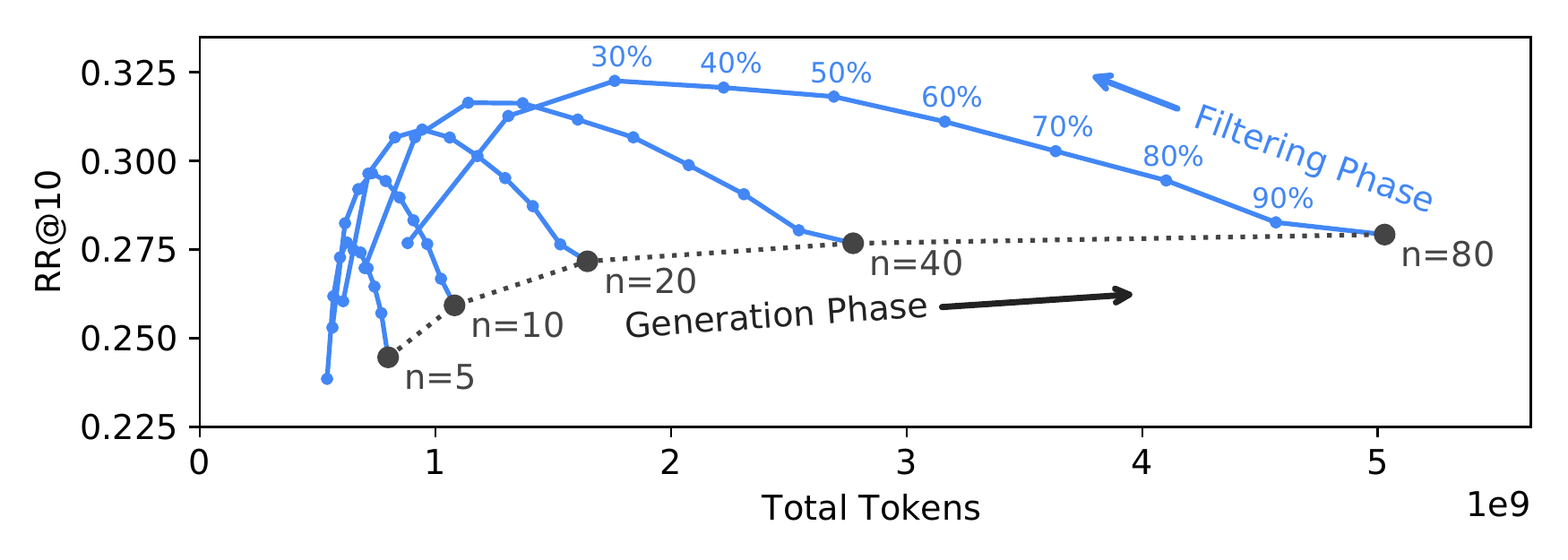}
    \vspace{-1em}
    \caption{Effectiveness (RR@10) on the Dev set, compared with the total number of indexed tokens. The generation phase is shown in dotted black (at various values of $n$), and the ELECTRA filtering phase is shown in solid blue (at various values of $p$).}
    \label{fig:lineplot}
\end{figure}

Digging a little deeper, Figure~\ref{fig:lineplot} shows the retrieval effectiveness of Doc2Query with various numbers of generated queries (in dotted black) and the corresponding performance when filtering using the top-performing ELECTRA scorer (in solid blue). We observe that performing relevance filtering at each value of $n$ improves the retrieval effectiveness. For instance, keeping only 30\% of expansion queries at $n=80$, performance is increased from 0.279 to 0.323 -- a 16\% improvement.

In aggregate, results from Table~\ref{tab:main} and Figure~\ref{fig:lineplot} answer RQ1: Doc2Query\--\-- filtering can significantly improve the retrieval effectiveness of Doc2Query across various scoring models, numbers of generated queries ($n$) and thresholds ($p$).

\begin{table}[b!]
\centering
\caption{Retrieval effectiveness comparison for comparable indexing computational budgets (in hours of GPU time). Values of $n$ without a filter are chosen to best approximate the total compute hours or the Dev effectiveness of the corresponding filtered version. Significant differences between in RR@10 performance are indicated with * (paired t-test, $p<0.05$).}
\label{tab:indexing}
\begin{tabular}{llrrrl}
&&\multicolumn{1}{c}{GPU Hours} & \multicolumn{2}{c}{RR@10} \\
\cmidrule(lr){3-3}\cmidrule(lr){4-5}
$n$ & Filter & Gen+Filt=Tot & Dev & Dev2 & Comment \\
\toprule

5 & ELECTRA & 20 + 15 = \phantom{0}34 &\bf 0.273 &\bf 0.270 \\
11 & \textit{None} & 34 + \phantom{0}0 = \phantom{0}34 & *0.261 & *0.256 & \textit{$-$4\% Dev RR for sim. GPU hrs} \\
31 & \textit{None} & 99 + \phantom{0}0 = \phantom{0}99 &\bf 0.273 & 0.265 & \textit{$\times 2.9$ GPU hrs to match Dev RR} \\
\midrule
10 & ELECTRA & 32 + 25 = \phantom{0}57 &\bf 0.292 &\bf 0.292 \\
18 & \textit{None} & 59 + \phantom{0}0 = \phantom{0}59 & *0.270 & *0.260 & \textit{$-$8\% Dev RR for sim. GPU hrs} \\
\midrule
20 & ELECTRA & 66 + 47 = 113 &\bf 0.307 &\bf 0.303 \\
36 & \textit{None} & 113 + \phantom{0}0 = 113 & *0.275 & *0.265 & \textit{$-$10\% Dev RR for sim. GPU hrs} \\
\midrule
40 & ELECTRA & 128 + 86 = 214 &\bf 0.316 &\bf 0.310 \\
68 & \textit{None} & 216 + \phantom{0}0 = 216 & *0.279 & *0.267 & \textit{$-$12\% Dev RR for sim. GPU hrs} \\

\bottomrule
\end{tabular}
\end{table}

Next, we explore the trade-offs in terms of effectiveness, efficiency, and storage when using Doc2Query\--\--. Table~\ref{tab:main} includes the mean response time and index sizes for each of the settings. As expected, filtering reduces the index size since fewer terms are stored. For the best-performing setting ($n=80$ with ELECTRA filter), this amounts to a 33\% reduction in index size (1.41 GB down to 0.95~GB). Naturally, such a reduction has an impact on query processing time as well; it yields a 23\% reduction in mean response time (30ms down to 23ms).

\looseness -1  Doc2Query\--\-- filtering adds substantial cost an indexing time, mostly due to scoring each of the generated queries. Table~\ref{tab:indexing} \crcedit{reports} the cost (in hours of GPU time) of the generation and filtering phases. We observe that ELECTRA filtering can yield up to a 78\% increase in GPU time ($n=10$). However, we \crcedit{find} that the improved effectiveness makes up for this cost. To demonstrate this, we allocate the time spent filtering to generating \crcedit{additional} queries for each passage. For instance, the 15 hours spent scoring $n=5$ queries could instead be spent generating 6 more queries per passage (for a total of $n=11$). We find that when comparing against an unfiltered $n$ \crcedit{that} closely approximates the total time when filtering, the filtered results consistently yield significantly higher retrieval effectiveness. \crcedit{As the computational budget increases, so does the margin between Doc2Query and Doc2Query\--\--, from 4\% at 34 hours up to 12\% at 216 hours.}

\crcedit{From the opposite perspective, Doc2Query consumes 2.9$\times$ or more GPU time than Doc2Query\--\-- to achieve similar effectiveness ($n=13$ with no filter vs. $n=5$ with ELECTRA filter). Since the effectiveness of Doc2Query flattens out between $n=40$ and $n=80$ (as seen in Figure~\ref{fig:lineplot}), it likely requires a massive amount of additional compute to reach the effectiveness of Doc2Query\--\-- at $n\geq10$, if that effectiveness is achievable at all. These comparisons show that if a deployment is targeting a certain level of effectiveness (rather than a target compute budget), Doc2Query\--\-- is also preferable to Doc2Query.}

These results collectively answer RQ2: Doc2Query\--\-- provides higher effectiveness at lower query-time costs, even when controlling for the additional compute required at index time.

\section{Conclusions}\label{s:conc}

This work demonstrated that there are untapped advantages in generating natural-language for document expansion. Specifically, we presented Doc2Query\--\--, which is a new approach for improving the effectiveness and efficiency of the Doc2Query model by filtering out the least relevant queries. We observed that a 16\% improvement in retrieval effectiveness can be achieved, while reducing the index size by 33\% and mean query execution time by 23\%.

\crcedit{The technique of filtering text generated from language models using relevance scoring is ripe for future work. For instance, relevance filtering could potentially apply to approaches that generate alternative forms of queries~\cite{Yu2020FewShotGC}, training data~\cite{bonifacio2022inpars}, or natural language responses to queries~\cite{Das2019MultistepRI} --- all of which are potentially affected by hallucinated content. Furthermore, future work could explore approaches for relevance filtering over masked language modelling expansion~\cite{epic}, rather than sequence-to-sequence expansion.}

\section*{Acknowledgements}

Sean MacAvaney and Craig Macdonald acknowledge EPSRC grant EP/R018634/1: Closed-Loop Data Science for Complex, Computationally- \& Data-Intensive Analytics.

\bibliographystyle{splncs04}
\bibliography{biblio}

\end{document}

%% file: fig_ex_msm.tex
\begin{figure}
\centering
\scalebox{0.78}{
\begin{tabular}{|p{3in}|p{3in}|}
\hline
\textbf{Original Passage:} Barley (Hordeum vulgare L.), a member of the grass family, is a major cereal grain. It was one of the first cultivated grains and is now grown widely. Barley grain is a staple in Tibetan cuisine and was eaten widely by peasants in Medieval Europe. Barley has also been used as animal fodder, as a source of fermentable material for beer and certain distilled beverages, and as a component of various health foods. &  \textbf{Generated Queries:}
\textcolor{rel1}{(1) where does barley originate from} $\cdot$
\textcolor{rel2}{(2) what is the name of the cereal grain used in tibetan cooking?} $\cdot$
\textcolor{rel3}{(3) what is barley used for} $\cdot$
\textcolor{rel1}{(1) what is barley in food} $\cdot$
\textcolor{rel0}{(0) what is bare wheat} $\cdot$
\textcolor{rel3}{(3) what family of organisms is barley in} $\cdot$
\textcolor{rel1}{(1) why is barley important in tibetan diet} $\cdot$
\textcolor{rel3}{(3) what is barley} $\cdot$
\textcolor{rel2}{(2) where is barley grown} $\cdot$
\textcolor{rel1}{(1) where was barley first grown and eaten} $\cdot$
\textcolor{rel1}{(1) where was barley first used} ...
\\
\hline
\end{tabular}
}
\caption{Example passage from MS MARCO and generated queries using the T5 Doc2Query model. The relevance of each query to the passage is scored by the authors on a scale of 0--3 using the TREC Deep Learning passage relevance criteria.}
\label{fig:ex_msm}
\end{figure}